\documentclass[runningheads]{llncs}
\usepackage{graphicx}
\usepackage{amsmath}
\usepackage{graphicx}
\usepackage{booktabs}
\usepackage{subfigure}
\usepackage{hyperref}

\begin{document}
\title{LSTM-RPA: A Simple but Effective Long Sequence Prediction Algorithm for Music Popularity Prediction\thanks{Supported by the National Natural Science Foundation of China under Grant Nos. 61806103, 61562068; ;Inner Mongolia discipline inspection and supervision big data laboratory open project fund No. imdbd2020013;Science and technology planning project of Inner Mongolia Autonomous Region No. jh20180175;Science and technology research project of colleges and universities in Inner Mongolia Autonomous Region No. njzy21578, No. njzy21551; the Youth Innovation and Entrepreneurship Talents of Inner Mongolia “grassland talents” Project; the Graduate Innovation Fund of Inner Mongolia Normal University under Grant No. CXJJS18112;} \thanks{This work has been published on CNKI and the paper has been published in \emph{Journal of Computer Engineering and Applications}. \url{https://kns.cnki.net/kcms/detail/11.2127.TP.20210915.1447.010.html}}}
\author{Kun Li\inst{1} \and Meng Li \inst{2} \and
Yanling Li\inst{3} \and Min Lin \inst{4}}
\authorrunning{Kun et al.}
\institute{Inner Mongolia Normal University, 81 Zhaowuda Road, Saihan District, Hohhot, Inner Mongolia, China \\
\email{\{cisha1573,limeng20131015\}@163.com,\{cieclyl,linmin\}@imnu.edu.cn}}
\maketitle              % typeset the header of the contribution
\begin{abstract}
The big data about music history contains information about time and users' behavior. Researchers could predict the trend of popular songs accurately by analyzing this data. The traditional trend prediction models can better predict the short trend than the long trend. In this paper, we proposed the improved LSTM Rolling Prediction Algorithm (LSTM-RPA), which combines LSTM historical input with current prediction results as model input for next time prediction. Meanwhile, this algorithm converts the long trend prediction task into multiple short trend prediction tasks. The evaluation results show that the LSTM-RPA model increased F score by 13.03\%, 16.74\%, 11.91\%, 18.52\%, compared with LSTM, BiLSTM, GRU and RNN. And our method outperforms tradi-tional sequence models, which are ARIMA and SMA, by 10.67\% and 3.43\% improvement in F score.Code:\url{https://github.com/maliaosaide/lstm-rpa}

\end{abstract}
\section{Introduction}
With the development of network technology, music as one of the streaming media has also been developed rapidly, such as QQ Music \footnote{https://y.qq.com/} , Spotify\footnote{https://www.spotify.com/} , Ali music \footnote{https://www.xiami.com/}  and NetEase Cloud Music \footnote{https://music.163.com/}. People could listen, share, download, collect music, and communicate their music ideas online. At the same time, a large number of users' behavior data are generated. The data from different users have implicit relationships, although they seem like irrelevant. And it also attracted the attention of many researchers and Internet companies. With the help of music popularity prediction, enterprises can formulate accurate propaganda strategies. The researchers recommend songs for users accurately by analyzing the active users' music play sequence[1], musical complexity (chroma, rhythm, timbre, and arousal)[2], and Twitter users’ music-listening behaviors[3]. Meanwhile, some studies analyzed the music semantics constructs (genre, mood, instrumental, theme) [4], the music frequency spectrogram image [5], and the music raw audio signal [5] to predict the popularity of songs. We could predict the trend of music popularity precisely from the analysis of different aspects of music.

In this paper, we investigated the relationship between the historical information of audiences’ behavior and the trend of music popularity, and proposed an algorithm to improve the accuracy of long trend prediction. This historical information comes from the competition dataset of " Popular Music Prediction"\footnote{DataSet: https://tianchi.aliyun.com/competition/entrance/231531/information}-hosted by Tsinghua University and Alicloud Tianchi Big Data Platform. The goal of the contest is to predict music trends for the next month by analyzing this historical data. This competition attracted 5,475 teams all over the world, and this topic is of great research significance. We proposed the LSTM Rolling Prediction Algorithm (LSTM-RPA) model with different features by analyzing the time series of the data. This research could play an important role in business decision making. We believe that using historical behavioral information of audiences could improve the performance of trend prediction.

\section{Related Work}
From movie popularity prediction[6], stock market trend prediction[7], to short-trend traffic prediction[8], trend prediction and time series modeling have a long research history. Research [9] used the Auto Regressive Integrated Moving Average (ARIMA) model to predict the Dez dam reservoir inflow monthly, and their experimental results show that the Mean Square Error (MSE) of ARIMA is smaller than that of the Auto Regressive Moving Average (ARMA). In[10], the research found that the ARIMA model cannot predict the non-linear time series accurately. So, they proposed the hybrid model based on Artificial Neural Network (ANN) and Genetic Programming (GP), which predicted more accurately than ARIMA on non-linear time series.

Research [11] established the Artificial “Music Market” and found a successful song was not only related to the quality of the song itself but also to social preferences, which increased the difficulty of prediction. In[5], the researchers used the frequency spectrogram images of music raw audio signals as the CNN model input feature to predict popularity music, and the accuracy of this model is 61\% in the test set. Research [12] established HitMusicNet based on multimodal end-to-end Deep Learning (DL) architecture. HitMusicNet used autoencoder to compress high-dimensional features, which consist of audio, text lyrics, and meta-data to improve the accuracy of the model prediction. Research [13] divided the time features of music play-count into a basic trend and incremental trend, then they proposed Time Series based Music Prediction (TSMP) which is based on category optimal value selection. When songs’ play-count increased exponentially, the accuracy of the TSMP model prediction would be reduced. Therefore, they proposed an Extend-Time Series Music Prediction (E-TSMP) algorithm, which is combining Sub-Sequence Pattern Matching Method (SSPMM) and Additional Processing (AD), to improve the prediction accuracy. 

In DL, Recurrent Neural Network (RNN) could learn key information very well from sequences [14]. The Long Short-Term Memory (LSTM) network, the improvement of RNN, can solve the problem of RNN gradient explosion and gradient disappearance effectively in long sequences, and can transmit key information to further cells [15]. Thus, we used LSTM to predict songs’ play-count. We proposed a Rolling Prediction Algorithm (RPA) to solve the problem of low accuracy in the long-trend prediction task. In this paper, we build two kinds of models, that are traditional sequence models (ARIMA and Simple Moving Average (SMA)[16]) and DL sequence models (LSTM, Bidirectional Long-Short Term Memory (BiLSTM) [17], Gated Recurrent Unit (GRU) [18] and RNN model).

\section{Methodology}

\subsection{LSTM model}
Assume a time series  $X={x_{1},x_{2},x_{3},\dots,x_{n}},x_{1},x_{2},x_{3},\dots,x_{n}$ represent each point in time that contains the feature information. (1)$\sim$(6) shows the LSTM formula:
\begin{align}
&f_{t}=\sigma(W_{f}\cdot h_{t-1}+W_{f}\cdot x_{t}+b_{f})\\
&i_{t}=\sigma(W_{i}\cdot h_{t-1}+W_{i}\cdot x_{t}+b_{i})\\
&\tilde{C}=\tanh(W_{c}\cdot h_{t-1}+W_c\cdot x_{t}+b_{c})\\
&C_{t}=f_{t}\cdot C_{t-1}+i_{t}\cdot\tilde{C}\\
&o_{t}=\sigma(W_{o}\cdot h_{t-1}+W_{o}\cdot x_{t}+b_{o})\\
&h_{t}=o_{t}\cdot\tanh(C_{t})
\end{align}
In LSTM block,$\sigma$ is the sigmoid function, $\tilde{C}$ is the new memory of the current layer. In time t, $C_{t}$ is the memory cell block, $h_{t}$ is the hidden state.

\subsection{Single Feature LSTM Model}
\begin{figure}[!ht]
\centering
\includegraphics[width=\textwidth]{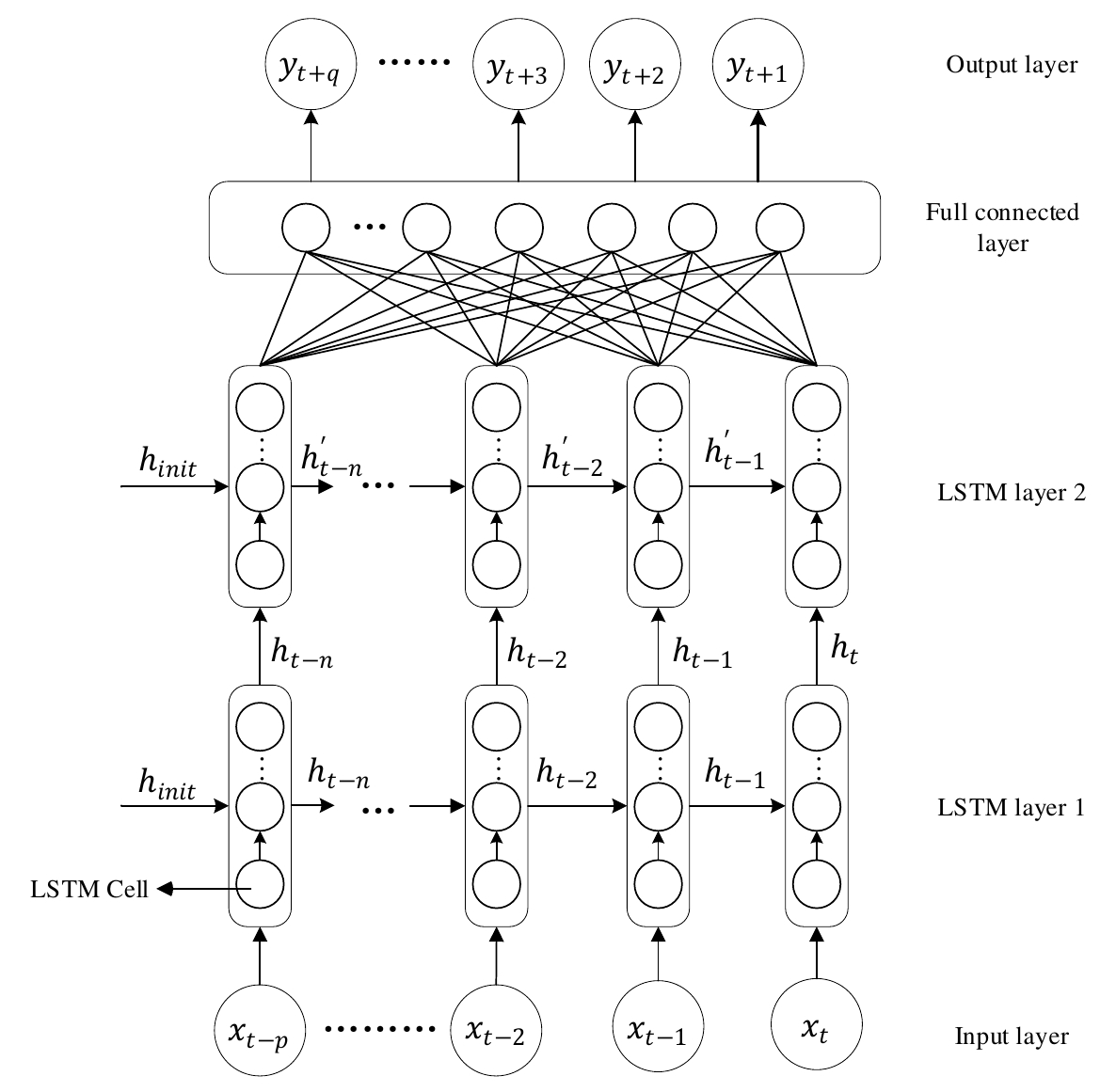}
\caption{Single Feature LSTM Model.} \label{fig:1}
\end{figure}

In the music popularity prediction task, one of the most important features is the songs’ historical play-count. For this single feature, we built the Single Feature LSTM Model (SF-LSTM) as shown in Fig.~\ref{fig:1}. This model inputs time series data of different time steps, then these data respective go through two LSTM layers which consist of 64 and 32 neurons respectively. We use ReLU as an activation function to reduce information loss. After that, The SF-LSTM uses a full-connected layer to reduce dimension, and outputs the prediction results. $p$ is the input time step, $q$ is the output time step, and $ H_{init}$ is the initial hidden state.

\subsection{Multiple Features LSTM Model}
\begin{figure}[!ht]
\centering
\includegraphics[width=\textwidth]{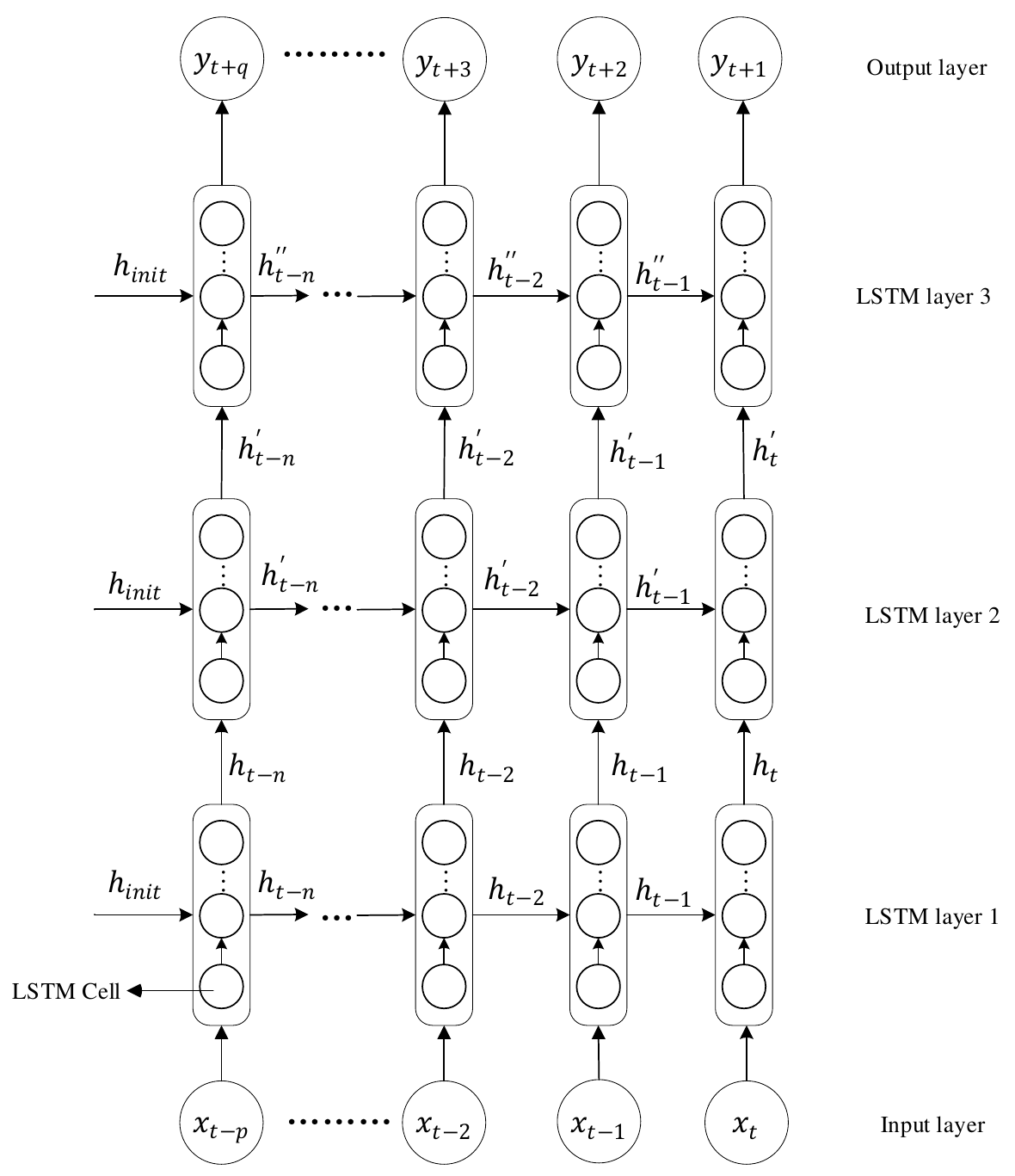}
\caption{Multiple Features LSTM Model.} \label{fig:2}
\end{figure}

Besides songs’ historical play-count, there are many other important features, such as weekly play-count, monthly play-count, download-count, collect-count, etc. We build the Multiple Features LSTM Model (MF-LSTM) as shown in Fig.~\ref{fig:2}, which consists of three LSTM layers. The first two layers are also LSTM layers that consist of 64 and 32 neurons respectively, ReLU as the activation function. But the last layer is the LSTM layer that consisted of q neurons with output step, sigmoid as the activation function. The particularity of model is that when the input step and the output step are equal, the input and output dimensions are same.

\subsection{Rolling Prediction Algorithm (RPA)}
Most of trend prediction algorithms only predict once[20]. That is, if we predict the next five days, only change the output layer into five neurons. In the case of mini-batch sets or lacking features, the model cannot get accuracy highly in long-trend prediction task. So, we proposed the Rolling Prediction Algorithm (RPA) which is based on moving average and Shortcut ideas [21] to solve this problem. This algorithm optimizes the long trend prediction for sequence models with mini-batch data sets or lacking features, and turn the long trend prediction task into multiple short trend prediction tasks.

\begin{figure}
\centering
\includegraphics[width=\textwidth]{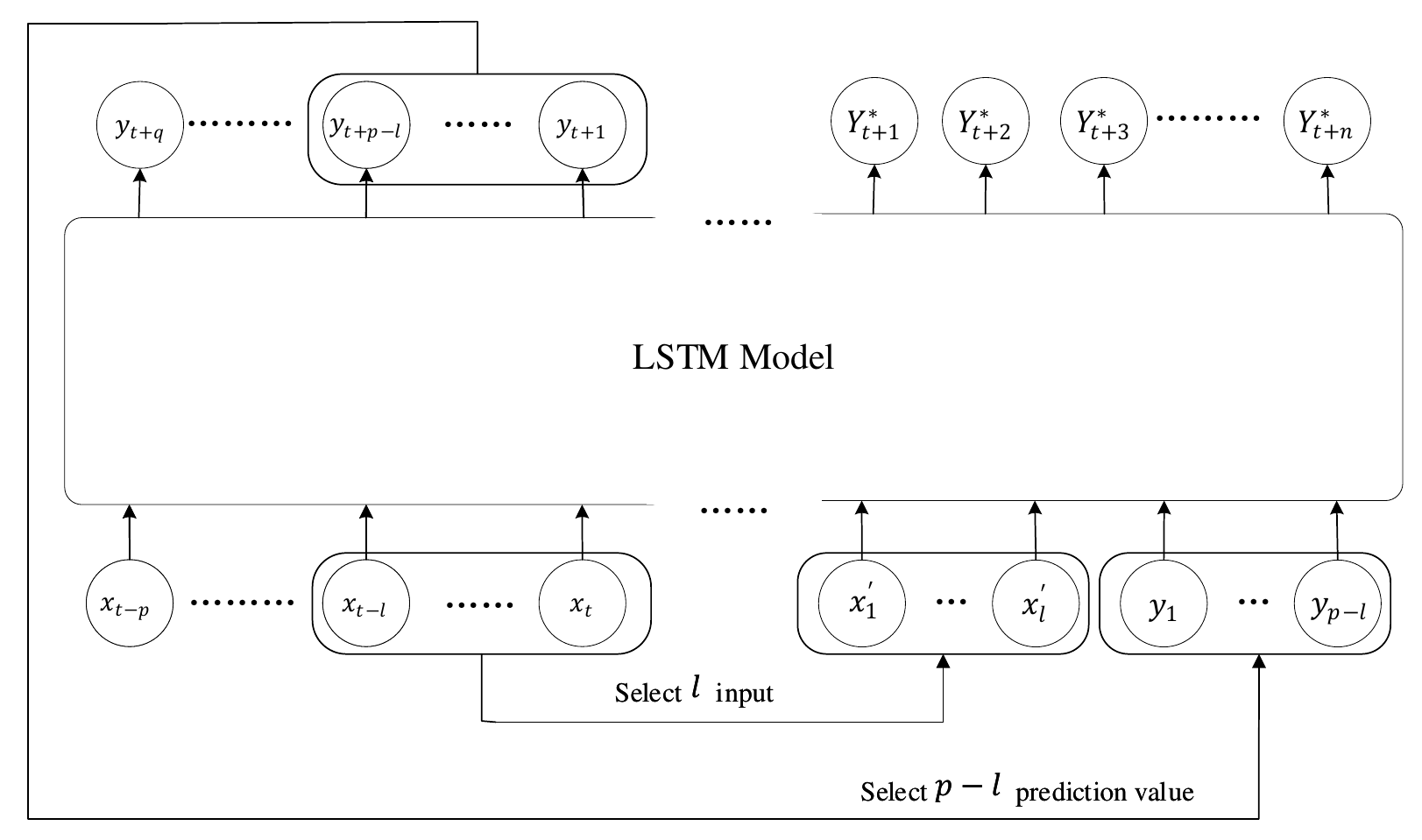}
\caption{LSTM-RPA model. We use SF-LSTM and MF-LSTM as $unit$ to make up of the LSTM-RPA model. $x_t,x_{t-1},\ldots,x_{t-p}$ is input for time $t-1$, and $y_{t+1},y_{t+2},\ldots,Y_{t+q}$ is prediction result for time $t$. At time $t+1$, The model prediction input $X_t^*$ consist of $X_{t-1}[x_1^{'},x_2^{'},\ldots,x_l^{'}]$ and $Y_t[y_1,y_2,\ldots,y_{p-l}]$ sequences, then the predicted result $Y_t^*[Y_{t+1}^*,Y_{t+2}^*,\ldots,Y_{t+n}^*]$ is acquired by the trained LSTM model.} \label{fig:3}
\end{figure}

The RPA considers transforming a sequence with time $T$ step into multiple prediction tasks with p length. When the RPA construct the input of the model in predicting time $t+1$, it brings in the time t-1 and t information together, so that the model can consider the information of $t-1$ and $t$ in predicting time $t+1$. (7)$\sim$(8) shows algorithm formula:
\begin{align}
&X_t^{*}=X_{t-1}[x_1^{'},x_2^{'},\ldots,x_l^{'}]+Y_t[y_1,y_2,\ldots,y_{p-l}]\\
&Y_{t+1}^{*}=unit(X_t^{*})
\end{align}
Set the model input sequence $X$, length $p$; output sequence $Y$, length $q$; and $l$ is rolling step, so $0\le l\le p,0\le p-l\le q$. In (7), $X_{t-1}$ is the input sequence for the model to predict $Y$ at time $t$, we reversely order select $l$ input values: $x_1^{'},x_2^{'},\ldots,x_l^{'}$. $Y_t$ is the model prediction result at the time $t$, we order select $p-l$ prediction values: $y_1,y_2,\ldots,y_{p-l}$. Above all, at the prediction $t+1$ time, the model input sequence $X_t^{*}$ consists of $X_{t-1}$ and $Y_t$. In (8), $unit$ is the trained sequence neural network, and the combination sequence $X_t^*$ gets the prediction result $Y_{t+1}^*$ by $unit$. We build the LSTM-RPA, the network structure is shown as Fig.~\ref{fig:3}.

\section{Experiments}
\subsection{Data preparation and analysis}
The dataset includes 183 days, 50 artists, 349946 users and 10842 songs, collected from March 1 to August 30, 2015. The sensitive information (Song\_id, Artist\_id, and User\_id) has been replaced by the fixed-length string, and one song belongs to only one artist. The data format shows in Table~\ref{tab:1},~\ref{tab:2}. 

\begin{table}
\centering
\caption{User action}\label{tab:1}
\begin{tabular*}{\textwidth}{@{\extracolsep{\fill}}ccccccc}
\hline\noalign{\smallskip}
&Name &Type &Description\\
\noalign{\smallskip}\hline\noalign{\smallskip}
&Song\_id &String &Track name\\
&Artist\_id &String &Artists’ name\\
&Ds &String &Date of data recording\\
&Gmt\_create &String &Time of user’s playing\\
&Action\_type &String &Type of songs’ action:1. play; 2. download; 3. collect\\
\noalign{\smallskip}\hline  
\end{tabular*}
\end{table}

\begin{table}
\centering
\caption{Relations between artists and songs}\label{tab:2}
\begin{tabular*}{\textwidth}{@{\extracolsep{\fill}}ccccccc}
\hline\noalign{\smallskip}
&Name &Type &Description\\
\noalign{\smallskip}\hline\noalign{\smallskip}
&Song\_id &String &Track name\\
&Artist\_id &String &Artists’ name\\
&Publish\_time &String &Publish time\\
&Gender	&String	&Gender: 1. Male; 2. Female; 3. Combination\\
&Language	&String	&Language:1. Chinese; 2.  French; 3. English\\
\noalign{\smallskip}\hline  
\end{tabular*}
\end{table}

Table~\ref{tab:1} shows users’ behavior, such as Track name, Artist, Date of data recording, Type of songs’ action. Table~\ref{tab:2} shows the relationship between the artists and the songs, such as the song belongs to which one artist, the song’s language, and the initial play-count. After preprocessing the data set, we found that one artist’s data was lost seriously. So we selected the other 49 artists’ data as the experimental data.
\subsection{Feature analysis and data slicing}
Considering the influence of the different features on model accuracy, we select daily songs’ play-count as the input feature in a single feature experiment, and select daily artist songs’ play-count, download-count and collect-count as the input feature in multiple features experiment.

At the same time, we chose the March to July dataset as the train and dev set divided by 122:31, and the August data as the test set. The experiments’ goal is to use the data from the first five months to predict daily songs’ play-count in August.

\begin{figure}[htbp]
\centering
\subfigure[A]{\includegraphics[width=5.5cm]{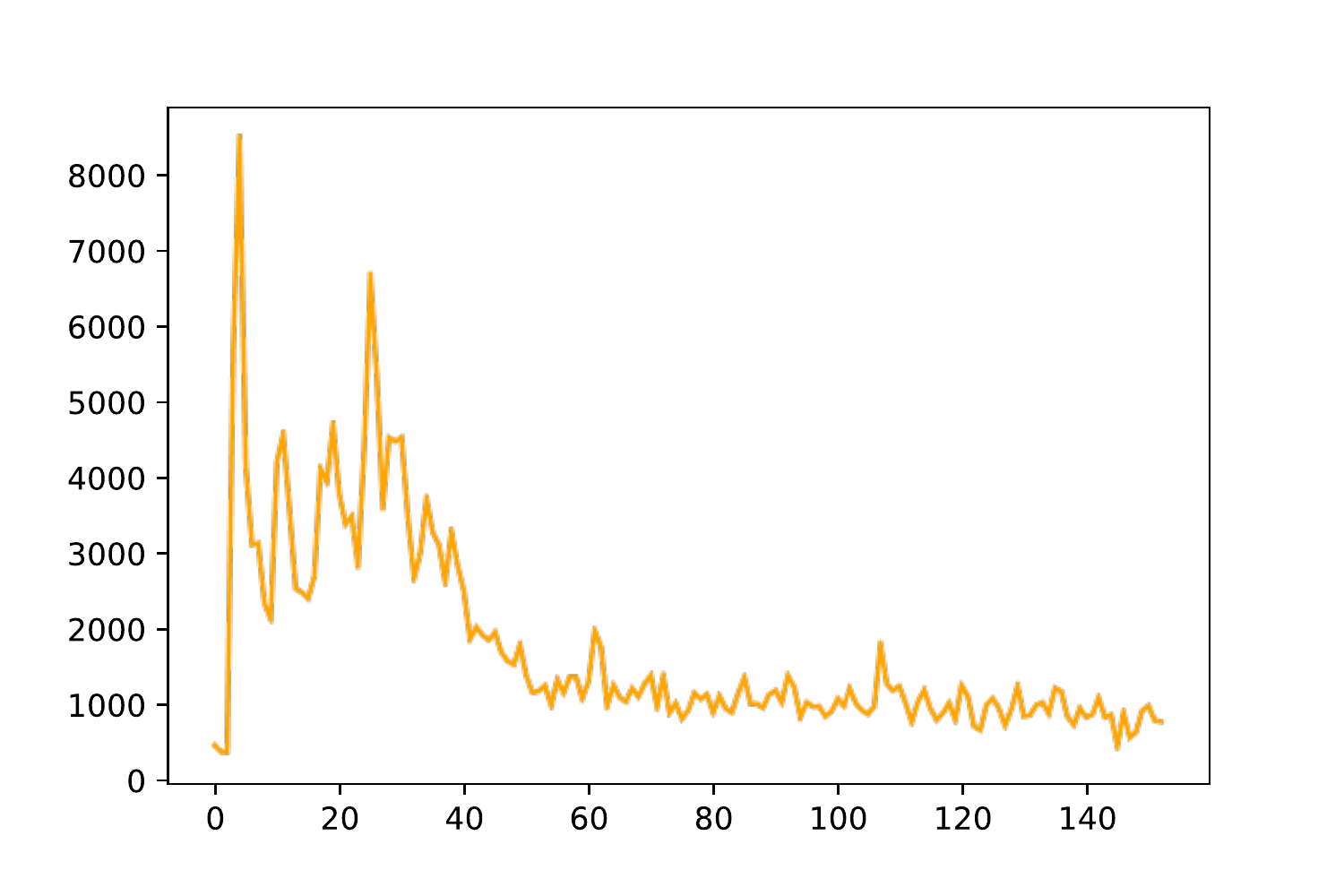}}
\quad
\subfigure[B]{\includegraphics[width=5.5cm]{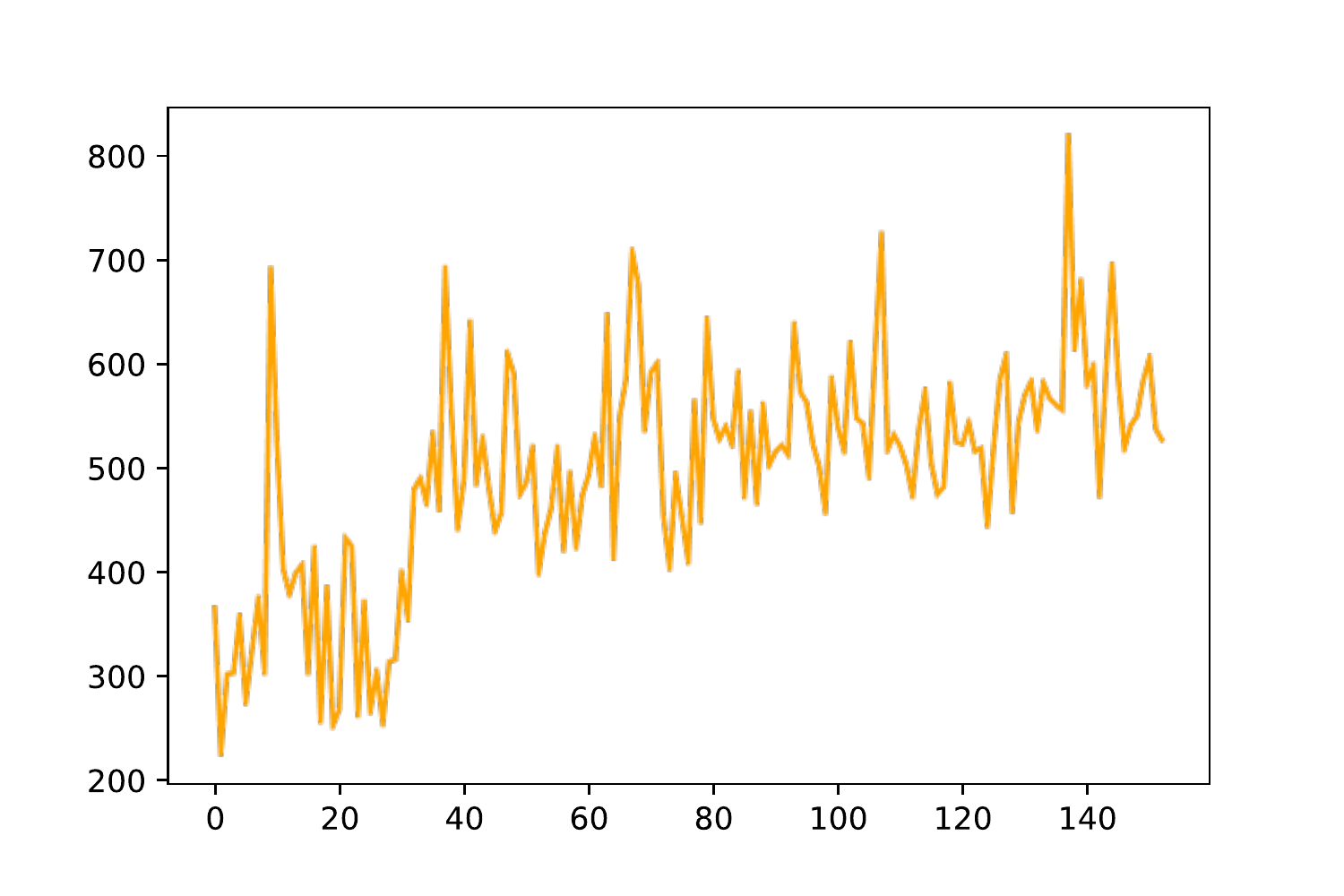}}
\quad
\subfigure[C]{\includegraphics[width=5.5cm]{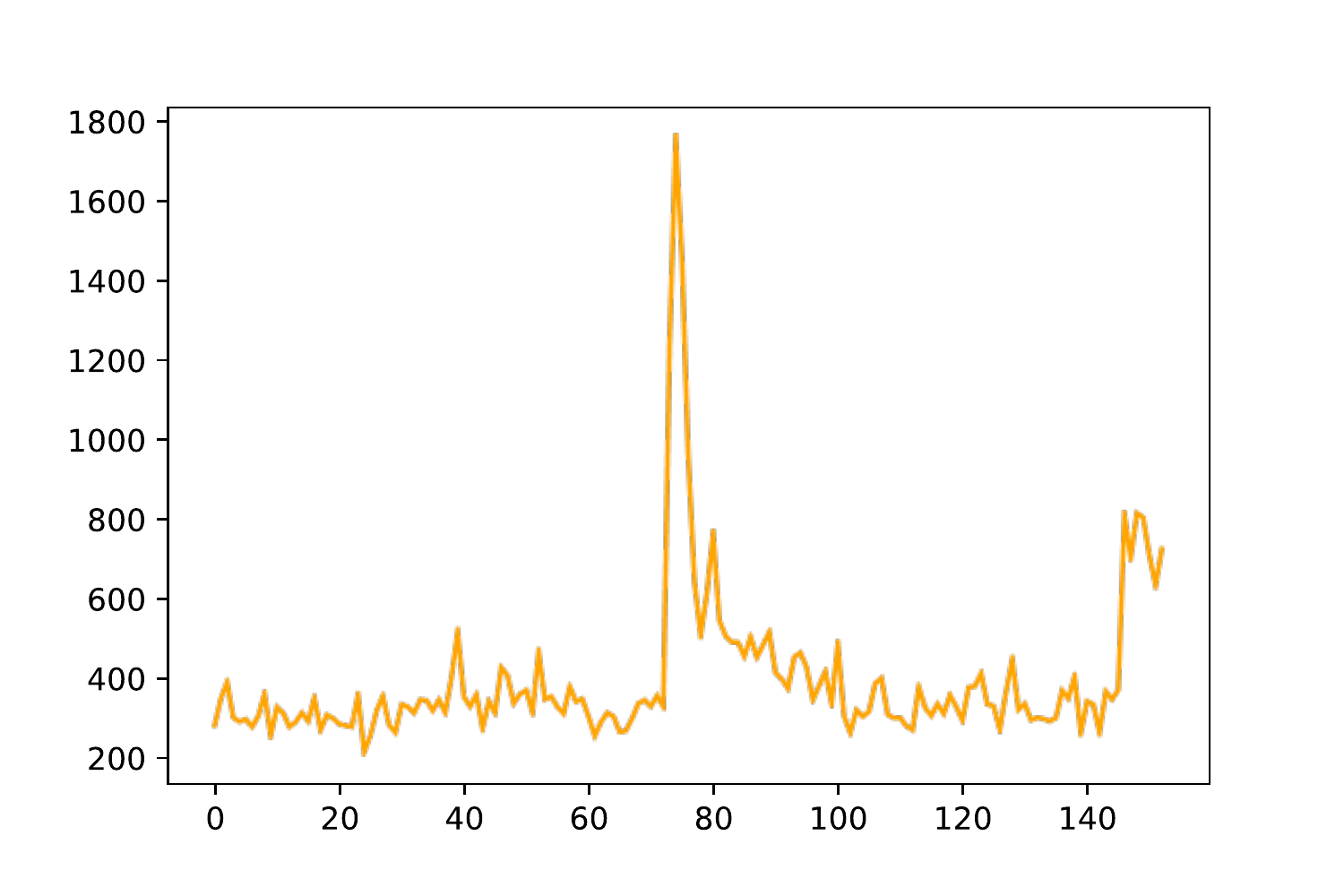}}
\quad
\subfigure[D]{\includegraphics[width=5.5cm]{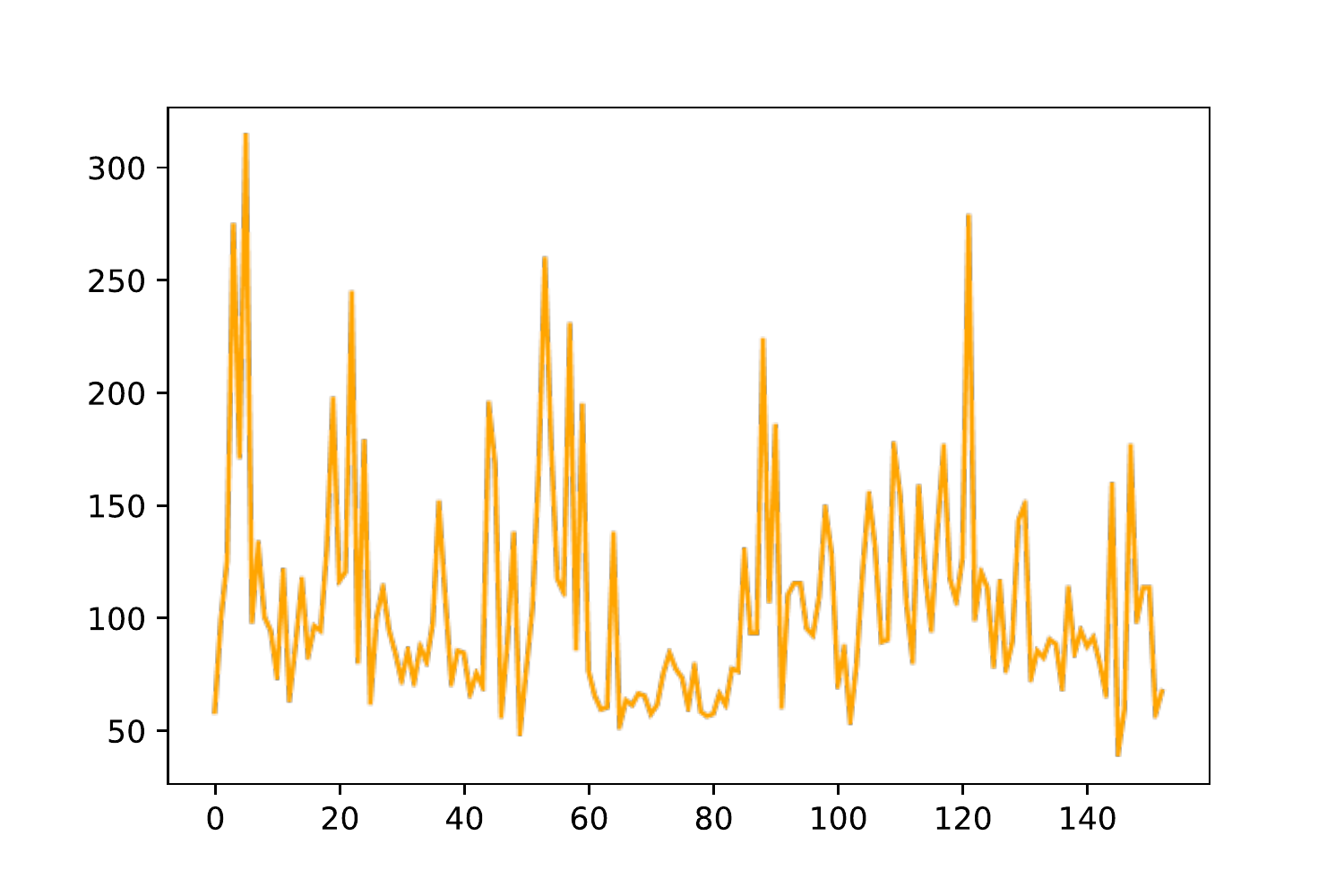}}
\caption{From March to July, the trend of songs playback by artists A, B, C, D. The x-axis denotes the date, its range is 1-122; the y-axis denotes the play-count}\label{fig:4}
\end{figure}

\subsection{Music popularity trend analysis}
Fig.~\ref{fig:4} shows the daily play-count trend of 4 artists, which selected randomly from 49 artists from March to July. The x-axis denotes the date, its range is 1-122. The y-axis denotes the play-count. Artist A's play-count fluctuated a lot in the 1 to 40 days, but after 40 days the play-count gradually stabilized around 1000 daily. It is speculated that before 40 days this fluctuation may be the recommended period after the artist releases a new album, or the season's top TV series, or movie songs produced by the artist. Artist B's total play-count trended to jitter upwards, with a rapid growth period of nearly 10 days after 30 days. This was possibly influenced by external platform recommendations or similar song recommendations. After 80 days, the play-count trended to fluctuate slowly. Artist C had a steady trend in 1~70 days, but on the 75th day there have huge fluctuations; we guess: (1) Error data. (2) Special holidays impact. (3) Influenced by sensational entertainment news. Artist D's daily play-count is extremely unstable.

Each artist's historical play-count data contains their unique time information, and there is a large gap between the different artists’ data. It is impossible to use one model to predict all artist songs’ play-count trend, so we build LSTM-RPA for 49 artists independently.

\subsection{Evaluation}
The competition organizer gives the evaluation function: F score, which is used to measure the difference between the predicted value and the actual value. a denotes artists $a\in W$, and $W$ is artists set. $Y_{a,d}$ is actual play-count on $d$ day, and $X_{a,d}$ is model prediction value on $d$ day. $\sigma_a$ is normalized variance $a$:
\begin{align}
\sigma_a=\sqrt{\frac{1}{N}\Sigma_{d=1}^N(\frac{X_{a,d}-Y_{a,d}}{Y_{a,d}})^2}
\end{align}
In (9), N is the total number of days predicted.  $\sigma_a$ indicate the gap between the $X_{a,d}$ and $Y_{a,d}$. If $\sigma_a$ is small, $(1-\sigma_a)$ is large which means the model’s prediction very accurate. $\phi_a$ is artist weight. 
\begin{align}
\phi_a=\sqrt{\Sigma_{d=1}^N Y_{a,d}}
\end{align}
Finally, the F score is defined.
\begin{align}
    F=\Sigma_{a\in W}(1-\sigma_a)\cdot\phi_a
\end{align}
In this paper, we use the F score as a measure of prediction accuracy. The larger the F score, the more accurate the prediction.

\subsection{Experimental result}
The music trend prediction is not only influenced by different features, but also by different rolling steps and time steps. When the input is single or multiple features, we researched the effect of different rolling and time steps on the accuracy of long trend prediction, and it is based on the Keras framework.
\subsection{Single feature LSTM-RPA experiment}
This experiment established and compared four kinds of neural networks: LSTM, BiLSTM, GRU, and RNN. The GRU and RNN network structure is the same as the SF-LSTM model. The BiLSTM model modifies the first two LSTM layers to bidirectional LSTM layers, which means data flows forward and backward only in the current bidirectional LSTM layer. The baseline result (black line) is the prediction score of the corresponding model without using the RPA. In this experiment, we only consider a positive score and change the negative score to 0.
\begin{figure}[h!]
\centering
\includegraphics[width=\textwidth]{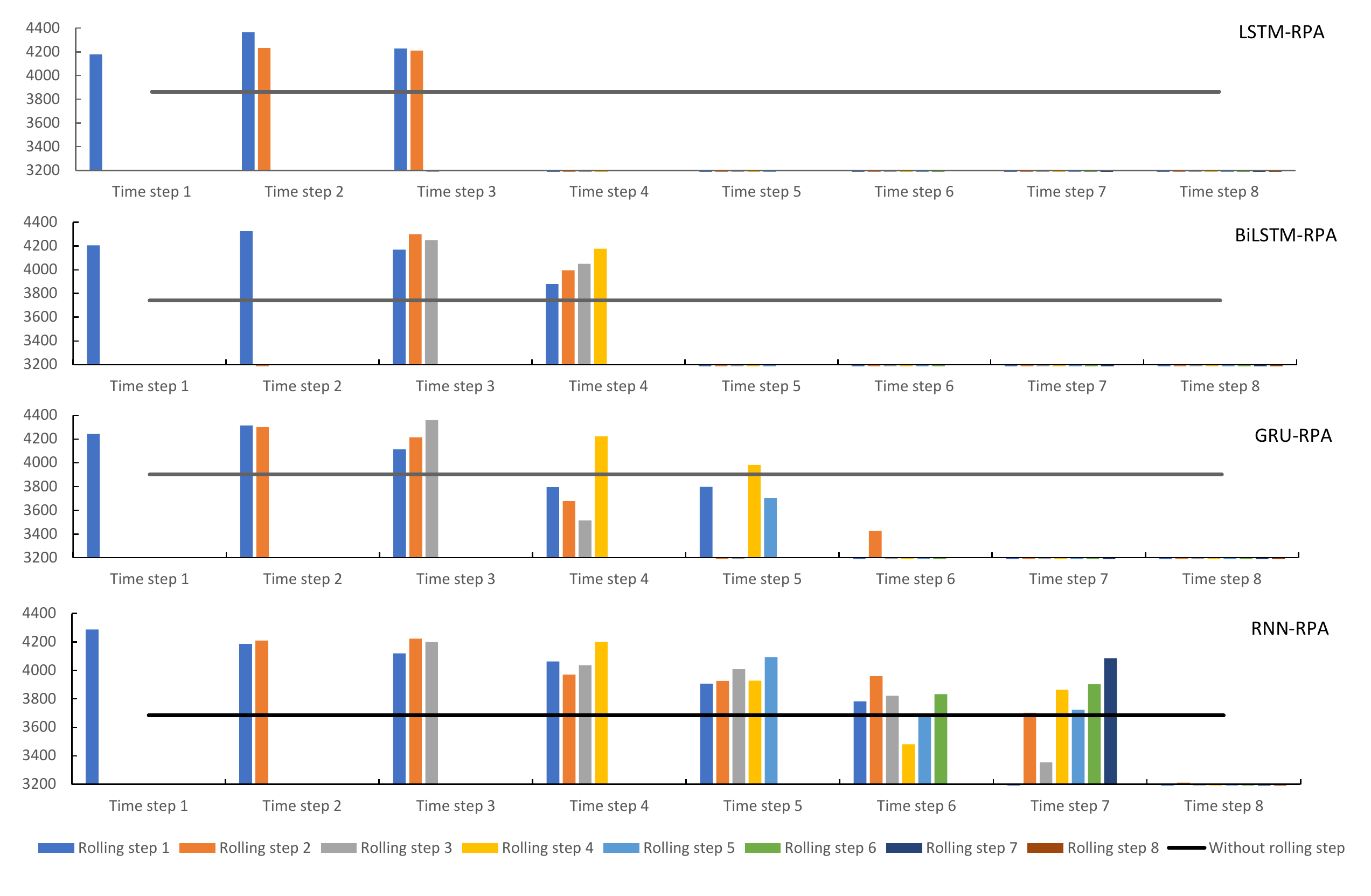}
\caption{F scores of different models with different time steps and rolling steps in a single feature experiment. The x-axis denotes the F score, the y-axis denotes the time step.}\label{fig:5}
\end{figure}
Fig.~\ref{fig:5} shows the predicted F score of 49 artists by different rolling prediction models under different time and rolling steps in the single feature. When the time step is less than 3, the LSTM-RPA prediction accuracy is better than the LSTM model (baseline). However, when the time step is greater than 3, the LSTM-RPA model prediction accuracy is greatly reduced due to the overfitting. In first four time steps, the BiLSTM-RPA model has an underfitting (F score is 276.9514) at the time step 2, and other prediction scores are higher than the baseline score. At the time step 4, different rolling steps have different effects for the model’s prediction accuracy. When rolling step increase, the model’s accuracy is also improved. As a variant of the LSTM-RPA, the GRU-RPA model also has a high prediction accuracy when the time step is less than 3. However, different from the LSTM-RPA model, the GRU-RPA model still has prediction accuracy when the time step is greater than 3. More than the half time-step, the RNN-RPA model’s prediction score is greater than the baseline score.

\subsection{Multiple features LSTM-RPA experiment}
Because songs’ play-count trend is not only influenced by historical data but also influenced by songs’ download-count and collect-count. Based on the single feature LSTM-RPA experiment, we modified four neural networks structure. GRU and RNN network structures are same as the MF-LSTM. The output layer of BiLSTM model is the LSTM layer and others are the bidirectional LSTM layer. The black line is the baseline score.
\begin{figure}[h!]
\centering
\includegraphics[width=\textwidth]{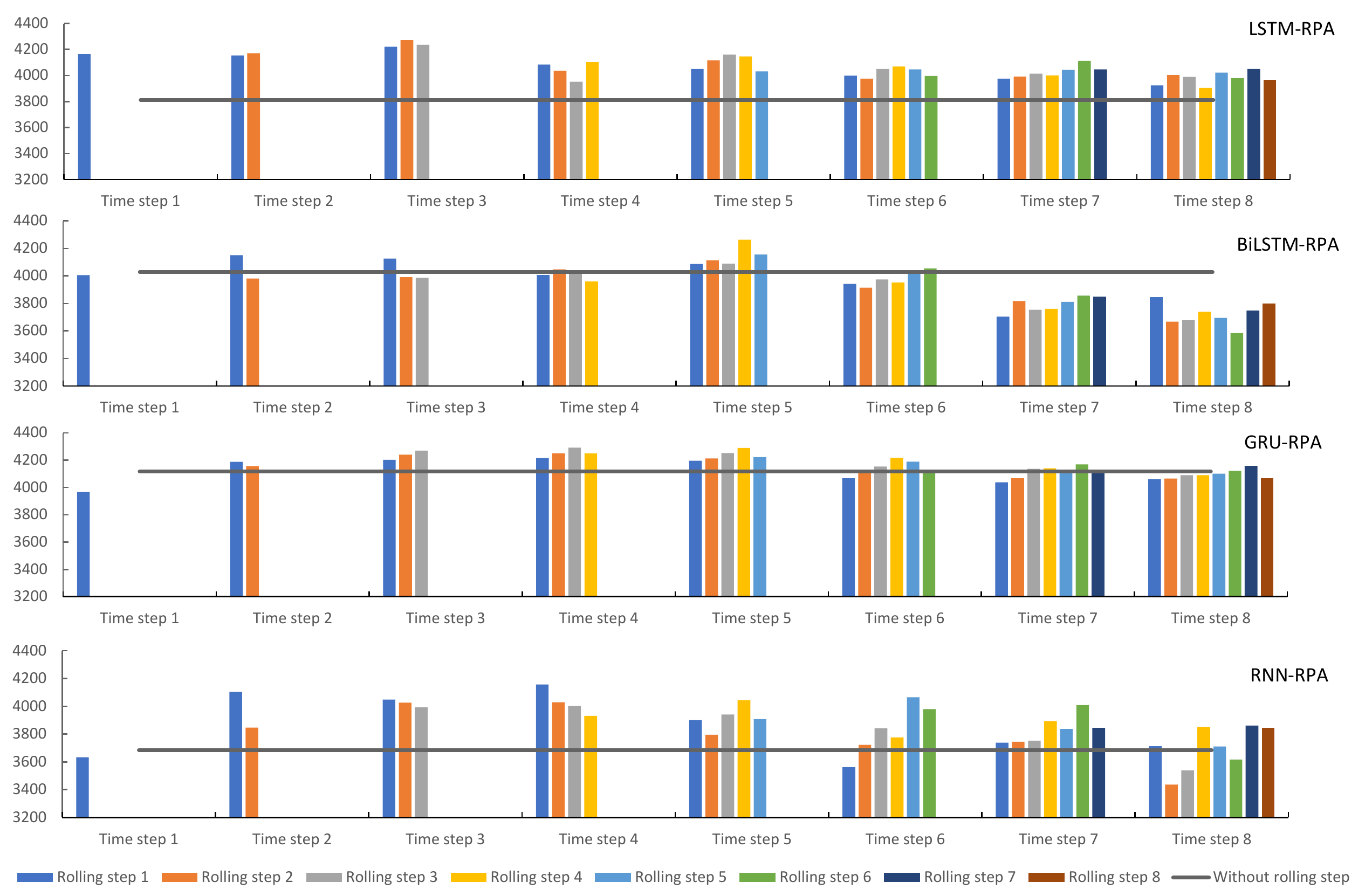}
\caption{F scores of different models with different time step and rolling step in the multi-feature experiment. The x-axis denotes the F score, the y-axis denotes the time step.}\label{fig:6}
\end{figure}
Fig.~\ref{fig:6} shows the F score of different models with different time and rolling steps in the multiple feature experiment. When increasing the features, the LSTM-RPA model prediction accuracy is better than the LSTM model without RPA at all times steps. When the time step is greater than 5, the BiLSTM-RPA model prediction accuracy decreases with the increase of time step. This is because that the amount of training data increases when the time step increases. At the same time, the amount of forward and reverse training data increases in the bidirectional network structure, which makes the model strengthen the memory of some non-necessary data, and thus leads to the decrease of the prediction accuracy. In the GRU-RPA model experiment, the RPA could optimize the baseline which has best score in all model’s baselines. The prediction accuracy of the RNN-RPA model is greatly influenced by different rolling steps in the case of the same time step, but some prediction F score is still higher than the baseline.

\subsection{Experimental analysis}
The experiment explored the effects of single and multiple features on the RPA, then selected and compared each RPA model’s best F scores and corresponding baseline under different features. Table~\ref{tab:3} shows the best F scores and baseline of each RPA model in the single feature experiment. The LSTM-RPA model has the highest F score and the lowest average error among the four models. And the F score of GRU-RPA and LSTM-RPA model are similar. The RPA could improve the prediction accuracy of RNN model better than other models. This experiment shows that the RPA has more than 10\% optimization effect for these sequence models in the single feature experiment.
\begin{table}
\centering
\caption{The F score and baseline of each rolling prediction model in the single feature experiment}\label{tab:3}
\begin{tabular*}{\textwidth}{@{\extracolsep{\fill}}ccccccc}
\hline\noalign{\smallskip}
&Model using RPA &Best F score &Baseline score &Optimum ratio &Average error\\
\noalign{\smallskip}\hline\noalign{\smallskip}
&LSTM &\textbf{4366.81} &3863.12 &13.03\% &\textbf{26.34}\\
&BiLSTM &4326.58 &3740.34 &15.65\% &27.19\\
&GRU &4360.07 &\textbf{3902.00} &11.73\% &26.49\\
&RNN &4287.89 &3684.44 &\textbf{16.37\%} &27.96\\
\noalign{\smallskip}\hline  
\end{tabular*}
\end{table}

\noindent Table~\ref{tab:4} shows the best F scores and baseline of different rolling prediction models in the multi-feature experiment. The LSTM-RPA, BiLSTM-RPA, and GRU-RPA have little difference in the best F score. And compared with their baselines, the RPA has the highest improvement effect on the LSTM-RPA model. But this optimization effect less than the RNN-RPA model. In all models, the GRU-RPA model has the highest F score and the lowest average error. These experimental results show that the RPA still has more than a 4\% optimization effect in the multi-feature experiment.

\begin{table}[!t]
\centering
\caption{The F score and baseline of each rolling prediction model in the multi-feature experiment}\label{tab:4}
\begin{tabular*}{\textwidth}{@{\extracolsep{\fill}}ccccccc}
\hline\noalign{\smallskip}
&Model using RPA &Best F score &Baseline score &Optimum ratio &Average error\\
\noalign{\smallskip}\hline\noalign{\smallskip}
&LSTM &4273.16 &3813.11 &12.06\% &28.26\\
&BiLSTM &4263.43 &4027.17 &5.89\% &28.46\\
&GRU &\textbf{4291.26} &\textbf{4117.23} &4.22\% &\textbf{27.89}\\
&RNN &4198.39 &3685.90 &\textbf{12.81}\% &30.60\\
\noalign{\smallskip}\hline  
\end{tabular*}
\end{table}

Based on single-feature and multi-feature experiments, we selected the rolling prediction model with the highest predictive accuracy: single-feature LSTM-RPA, and compared with the baseline of different models and traditional trend prediction algorithms: ARIMA and SMA. Table~\ref{tab:5} shows the best F scores and average errors of traditional models and sequence models. The experimental results show that the LSTM-RPA model has a better F score than other models. The LSTM-RPA compared with ARIMA and SMA, F score increases by 10.67\% and 3.43\%, average error decreases by 32.64\% and 11.23\%, average error variance decreases by 368.730\% and 35.29\%. And it compared with LSTM, BiLSTM, GRU, and RNN, F score increases by 13.03\%, 16.74\%, 11.91\% and 18.52\%, average error decreases by 39.02\%, 48.55\%, 36.02\% and 52.88\%, average error variance decreases by 183.69\%, 252.83\%, 175.23\% and 225.79\%.

\begin{table}
\centering
\caption{The F scores and mean errors of best models}\label{tab:5}
\begin{tabular*}{\textwidth}{@{\extracolsep{\fill}}ccccccccccc}
\hline\noalign{\smallskip}
&Evaluation &ARIMA &SMA &LSTM-RPA &LSTM &BiLSTM &GRU &RNN\\
\noalign{\smallskip}\hline\noalign{\smallskip}
&F score &3945.53 &4221.99 &\textbf{4366.81} &3863.12 &3740.34 &3902.00 &3684.44\\
&Average error &34.94 &29.30 &\textbf{26.34} &36.62 &39.13 &35.83 &40.27\\
&Mean error variance &45.42 &13.11 &\textbf{9.69} &27.49 &34.19 &26.67 &31.57\\
\noalign{\smallskip}\hline  
\end{tabular*}
\end{table}
\noindent In summary, the model using RPA has higher prediction accuracy and lower average error than baseline in single-feature and multi-feature experiments. In the proportion of improvement of algorithm optimization, the RPA optimization effect on multi-feature experiments is lower than that on single-feature experiments. At the same time, the LSTM-RPA model has higher prediction accuracy than the traditional trend prediction models. And the LSTM-RPA model has a lower average error variance than the LSTM model. All in all, the RPA has some optimization effect in the long trend prediction task. 

\section{Conclusion}
In this paper, we analyzed the data about information of artist and user’s behavior, selected different features (play-count, download-count and collect-count) for experiments. The goal of experiments is to predict artists’ play-count daily for 30 days. We proposed the LSTM Rolling Prediction Algorithm (LSTM-RPA) based on the moving average and Shortcut idea to improve the accuracy of the model in long trend prediction task, by analyzing the relationship between the model’s historical input and current prediction results. At time t+1, the LSTM-RPA model can consider the information of time t-1 and t, and turn the long trend prediction task into multiple short trend prediction tasks. Experiment results show that the prediction accuracy of the single-feature LSTM-RPA model is better than the traditional model and DL sequence models. The LSTM-RPA compared with ARIMA and SMA, the F score increases by 10.67\% and 3.43\%, the average error decreases by 32.64\% and 11.23\%. And The LSTM-RPA increased F score by 13.03\%, 16.74\%, 11.91\% and 18.52\%, decreased average error by 39.02\%, 48.55\%, 36.02\%, and 52.88\%, compared with LSTM, BiLSTM, GRU, and RNN. So, our method could improve prediction accuracy of sequence model in long trend prediction task.

In the future, we will introduce the Self-Attention Mechanism to increase model’s accuracy in long trend prediction task.

\end{document}